\documentclass[runningheads]{llncs}
\usepackage{graphicx}
\usepackage{mathtools,xparse}
\usepackage{amssymb}
\usepackage[final]{pdfpages}
\usepackage{lipsum}
\usepackage{array}
\usepackage{multirow}

\begin{document}
\title{Neural~Named~Entity~Recognition~for~Kazakh}

%
%\titlerunning{Abbreviated paper title}
% If the paper title is too long for the running head, you can set
% an abbreviated paper title here
%
\author{Gulmira Tolegen\and
Alymzhan Toleu\and
Orken Mamyrbayev \and \\
Rustam Mussabayev}

\institute{Institute of Information and Computational Technologies\\
Almaty, Kazakhstan\\
\email{\{gulmira.tolegen.cs\}@gmail.com,\{alymzhan.toleu\}@gmail.com}}

% \author{Gulmira Tolegen\inst{1}\orcidID{0000-1111-2222-3333} \and
% Alymzhan Toleu\inst{2,3}\orcidID{1111-2222-3333-4444} \and
% Third Author\inst{3}\orcidID{2222--3333-4444-5555}}

%
\authorrunning{Tolegen Gulmira et al.}
% First names are abbreviated in the running head.
% If there are more than two authors, 'et al.' is used.
%
%\institute{Princeton University, Princeton NJ 08544, USA \and
%Springer Heidelberg, Tiergartenstr. 17, 69121 Heidelberg, Germany
%\email{lncs@springer.com}\\
%\url{http://www.springer.com/gp/computer-science/lncs} \and
%ABC Institute, Rupert-Karls-University Heidelberg, Heidelberg, Germany\\
%\email{\{abc,lncs\}@uni-heidelberg.de}}
%
\maketitle              % typeset the header of the contribution

\begin{abstract}
We present several neural networks to address the task of named entity recognition for morphologically complex languages (MCL).
Kazakh is a morphologically complex language in which each root/stem can produce hundreds or thousands of variant word forms.
This nature of the language could lead to a serious data sparsity problem, which may prevent the deep learning models from being well trained for under-resourced MCLs.
In order to model the MCLs' words effectively, we introduce root and entity tag embedding plus tensor layer to the neural networks.
The effects of those are significant for improving NER model performance of MCLs. 
The proposed models outperform state-of-the-art including character-based approaches, and can be potentially applied to other morphologically complex languages.
\end{abstract}
\section{Introduction}
Named Entity Recognition (NER) is a vital part of information extraction.
It aims to locate and classify the named entities from unstructured text.
The different entity categories are usually the person, location and organization names, etc.
Kazakh language is an agglutinative language with complex morphological word structures.
Each root/stem in the language can produce hundreds or thousands of new words.
It leads to the severe problem of data sparsity when automatically identifying the entities.
In order to tackle the problem, Tolegen et al. (2016) \cite{Tolegen:16} have given the systematic study for Kazakh NER by using conditional random fields.
More specifically, the authors assembled and annotated the Kazakh NER corpus (KNC), and proposed a set of named entity features with the exploration of their effects.
To achieve a state-of-the-art result for Kazakh NER compared with other languages' NER.
Authors have manually designed feature templates, which in practice is a labor-intensive process and requires a lot of expertise. 
With the intention of alleviating the task-specific feature engineering, there has been increasing interest in using deep learning to solve the NER task for many languages.
However, the effectiveness of the deep learning for Kazakh NER is still unexplored. 
One of the aims of this work is to use deep learning for Kazakh NER to avoid the task-specific feature engineering and to achieve a new state-of-the-art result. 
As in similar studies\cite{Collobert:2011} the neural networks (NNs) produces high results for English or for other languages by using distributed word representations.
But using only surface word representation in deep learning is may not enough to reach the state-of-the-art results for under-resourced MCLs.
The main reason is that deep learning approaches are data hungry, their performance is strongly correlated with the amount of available training data.

In this paper, we introduce three types of representation for MCL including word, root and entity tag embeddings.
With the purpose of discovering how above embeddings contribute to model performance independently, we use a simple NN as the baseline to do the investigation.
We also improve this basic model from two perspectives.
One is to apply a tensor transformation layer to extract multi-dimensional interactions among those representations.
The other is to map each entity tag into a vector representation.
The result shows that the use of root embedding can lead to a significant improvement to the models in term of improving test results. 
Our NNs reached good outcomes by transferring intermediate representations learned on large unlabeled data.
We compare the NNs with the existing CRF-based NER system for Kazakh \cite{Tolegen:16} and the other bidirectional-LSTM-CRF \cite{N16-1030} that considered as the state-of-the-art in NER.
Our NNs outperforms the state-of-the-art and the result indicates that the proposed NNs can be potentially applied to other morphologically complex languages.

The rest of the paper is organized as follows: Section 2 reviews the existing work. 
Section 3 gives the named entity features used in this work.
Section 4 describes the details of neural networks.
Section 5 reports the results of experiments and the paper is concluded in Section 6 with future work.

\section{Related Work}
Named Entity Recognition have been studied for several decades, not only for English \cite{Chieu:2003:NER,Klein:2003:NER,konvens:17_tkachenko12o}, but also for other MCL, including Kazakh \cite{Tolegen:16} and Turkish \cite{Yeniterzi:2011:EMT,DBLP:conf/coling/SekerE12}. 
For instance, Chieu and Hwee Tou (2003) \cite{Chieu:2003:NER} presented a maximum entropy approach based NER systems for English and German, where the authors used both local and global features to enhance their models and achieved good performance in NER. 
In order to explore the flexibilities of the four diverse classifiers (Hidden Markov model, maximum entropy, transformation-based learning, robust linear classifier) for NER, the work \cite{Florian:2003:NER} showed that a combined system of these models under different conditions could reduce the F1-score error by a factor of 15 to 21\% on English data-set.
As known, the maximum entropy approach was suffering from the label bias problem \cite{Lafferty:2001}, then the researchers attempted to use CRF model \cite{McCallum:2003:NER} and presented CRF-based NER systems with a number of external features.
Such supervised NER systems were extremely sensitive to the selection of an appropriate feature set, in the work \cite{konvens:17_tkachenko12o}, the authors explored various combinations of a set of features (local and non-local knowledge features) and compared their impact on recognition performance for English.
Using the CRF with optimized feature template, they obtained a 91.02\% F1-score on the CoNLL 2003 \cite{TjongKimSang:2003} data-set.  

For Turkish, Yeniterzi (2011)\cite{Yeniterzi:2011:EMT} analyzed the effect of the morphological features, they utilized CRF that enhanced with several syntactic and contextual features, their model achieved an 88.94\% F1-score on Turkish test data.
In same direction Seker and Eryigit (2012)\cite{DBLP:conf/coling/SekerE12} presented a CRF-based NER system with their feature set, their final model achieved the highest F1-score (92\%). 
For Kazakh, Tolegen et al. (2016)\cite{Tolegen:16} annotated a Kazakh NER corpus (KNC), and carefully analyzed the effect of the morphological (6 features) and word type (4 features) features using CRF.
Their results showed that the model could be improved by using morphological features significantly, the final CRF-based NER system achieved an 89.81\% F1 on Kazakh test data. 
In this work, we use such CRF-based NER system as one baseline and make comparison to our deep learning models.
Recently, deep learning models including biLSTM have obtained a significant success on various natural languages processing tasks, such as POS tagging \cite{wieting-EtAl:2016:EMNLP2016,ling-EtAl:2015:EMNLP2,toleu-tolegen-makazhanov:2017:Short,8880244}, NER \cite{Chieu:2003:NER,kuru-can-yuret:2016:COLING}, machine translation \cite{DBLP:journals/corr/BahdanauCB14,He:NIPS:2016:6469}, word segmentation \cite{kuru-can-yuret:2016:COLING} and on other fields like speech recognition \cite{10.1080/23311916.2020.1727168,10.1080/23311916.2020.1727168,Graves06connectionisttemporal,orken2018,articleOrken:2018}.
As the state-of-the-art of NER,  in the study \cite{N16-1030}, the authors have explored various neural architectures for NER including the language independent character-based biLSTM-CRF models.
These type of models on German, Dutch and English have achieved 81.74\%, 85.75\% and 90.94\%.
Our models have several differences compared to other state-of-the-art.
One difference is that we introduce root embedding to tackle the problem of data sparsity caused by MCL.
The decoding part (refers it to CRF layer in literature \cite{N16-1030,P16-1101,W18-5605}) of NNs is combined into NNs using tag embedding.
Then the word, root and tag embeddings are efficiently incorporated and calculated by NNs in the same vector space, which allows us to extract higher-level vector features.

\begin{table}
\caption{The entity features, more details see Tolegen et al. \cite{Tolegen:16}}\label{table:statistics}
\centering

\begin{tabular}{clcl}
\hline
\multicolumn{2}{c}{Morphological features} & \multicolumn{2}{c}{Word type features}    \\ \hline
\multicolumn{2}{c}{Root}                   & \multicolumn{2}{c}{Case feature}          \\
\multicolumn{2}{c}{Part of speech}         & \multicolumn{2}{c}{Start of the sentence} \\
\multicolumn{2}{c}{Inflectional suffixes}  & \multicolumn{2}{c}{Latin spelling words}           \\
\multicolumn{2}{c}{Derivational suffixes}  & \multicolumn{2}{c}{Acronym}               \\
\multicolumn{2}{c}{Proper noun}            & \multicolumn{2}{c}{-}                     \\
\multicolumn{2}{c}{Kazakh Name suffixes}          & \multicolumn{2}{c}{-}                     \\ \hline
\end{tabular}
\end{table}

\section{Named Entity Features}
NER models are often enhanced with named entity features.
In this work, with the purpose of making a fair comparison, we utilize the same entity features proposed by Tolegen et al. (2016)\cite{Tolegen:16}.
The entity features are given in Table 1 with two categories: morphological and word type information.
Morphological features are extracted by using the morphological tagger of our implementation.
We used a single value (1 or 0) to represent each feature according to each word has the feature or not.
Then each word in the corpus contains an entity feature vector to feed into NNs with word, root and tag embeddings.

\section{The Neural Networks }
In this section, we describe our NNs for MCL NER. 
Unlike other NNs for English or other similar languages, we introduce three types of representations: word, root and tag embedding.
In order to explore the effect of root and tag embedding separately and clearly,  our first model is general deep neural network (DNN), which was first proposed by Bengio et al. (2003)\cite{Bengio:2003} for probabilistic language model, and re-introduced by Collobert et al. (2011)\cite{Collobert:2011} for multiple NLP tasks.
DNN also is a standard model for sequence labeling task and could be a strong baseline.
The second model is the extension of the DNN by applying a tensor layer to DNN.
The tensor layer can be viewed as a non-linear transformation that extracts higher dimensional interactions from the input.

The architecture of our NN is shown in Figure 1. 
The first layer is lookup table layer which extracts features for each word.
Here, the features are a window of words, and root ($S_i$) plus tag embedding ($t_{i-1}$).
The concatenation of these feature vectors are fed into the next several layers for feature extractions. 
The next layer is tensor layer and the remaining layers are standard NN layers.
The NN layers are  trained by backpropagation and the details of NNs are given in the following sections.

%\includepdf[width=70mm,height=30mm]{2019Ciciling.pdf}
\begin{figure}[!ht]
 \centering
 \includegraphics[width=8cm]{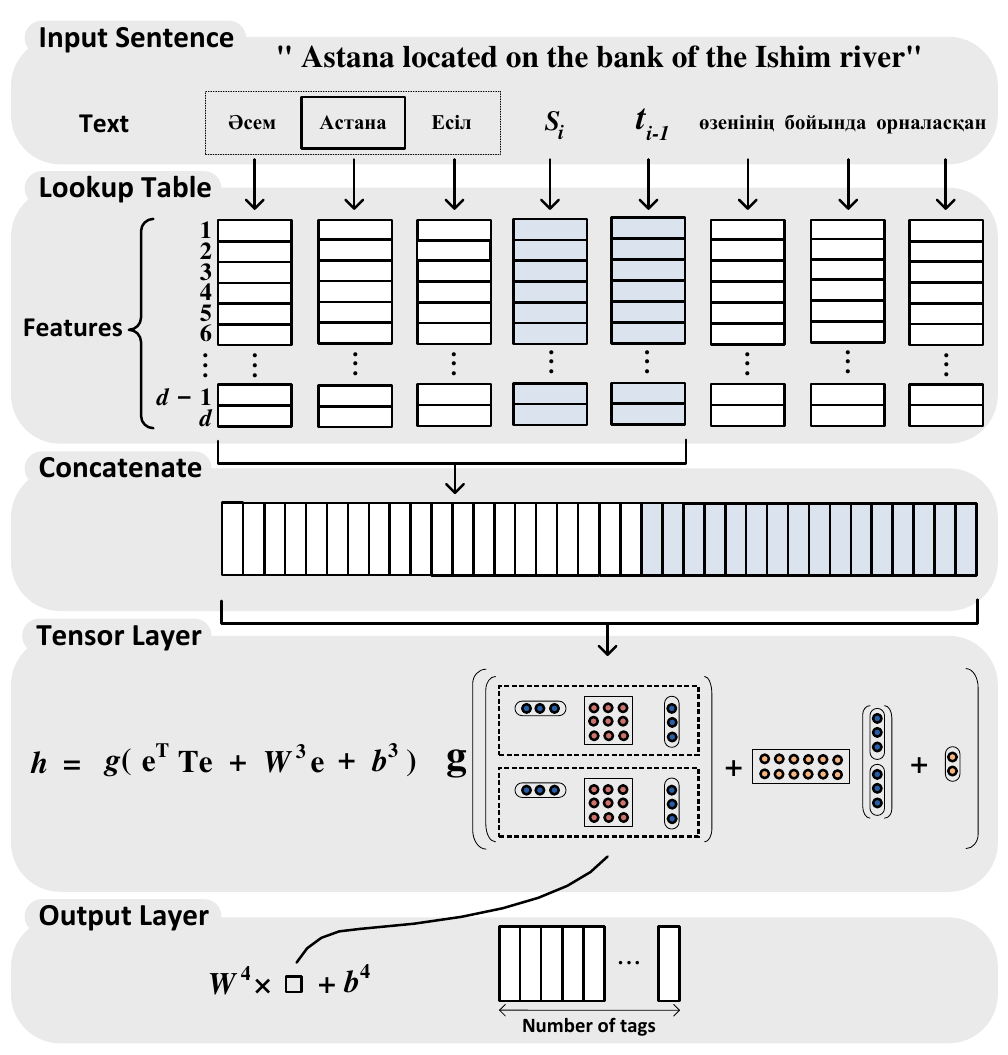}
 \caption{The architecture of the Neural Network.}
\end{figure}

\subsection{Mapping words and tags into feature vectors }
The NNs have two dictionaries\footnote{\scriptsize {The dictionary is extracted from training data and performed some pre-processing, namely lowercasing and word-stemming.
Words outside this dictionary are replaced by a single special symbol}.}: one for roots and another for words.
For simplicity, we will use one notation for both dictionaries in the following descriptions.
Let $\mathcal{D}$ be the finite dictionary, and for each word $x_i \in \mathcal{D}$ is represented as a $d$-dimensional vector $M_{x_i} \in \mathbb{R}^{1 \times d}$ where $d$ is word vector size (a hyper-parameter).
All word representation of the $\mathcal{D}$ are stored in a embedding matrix $M \in \mathbb{R}^{d \times |\mathcal{D}|}$ where $|\mathcal{D}|$ is size of the dictionary.
Each word $x_i \in \mathcal{D}$ corresponds to an index $k_i$ which is column index of the embedding matrix, and then the corresponding word embedding is retrieved by the lookup table layer $LT_M(\cdot)$:

\begin{equation}
\begin{aligned}
  LT_M(k_i) = M_{x_i}
\end{aligned}
\label{eq:1}
\end{equation}

Similar to word embedding, we introduce tag embedding $L \in \mathbb{R}^{d \times |\mathcal{T}|}$, where $d$ is the vector size and $\mathcal{T}$ is a tag set.
The lookup table layer can be seen as a simple projection layer where the word embedding for each context and tag embedding for the previous word is retrieved by lookup table operation.
To use these features effectively, we use a sliding window approach\footnote{\scriptsize {The words exceeding the sentence boundaries are mapped to one of two special symbols, namely ``start" and ``end" symbols.}}. 
More precisely, for each word $x_i \in X$, a window size word's embeddings are given by the lookup table layer:

\begin{equation}
\begin{aligned}
  f_{\theta}^{1}(x_i) =
  \begin{bmatrix}
  M_{x_{i-\frac {w}{2}}}
  \dots
  M_{x_i}
  \dots
  M_{x_{i+\frac {w}{2}}}
  ,
  S_{i}
  ,
  t_{i-1}
  \\
  \end{bmatrix}
\end{aligned}
\label{eq:1}
\end{equation}
where $f_{\theta}^{1}(x_i) \in \mathbb{R}^{1 \times wd}$ is $w$ word feature vectors, the $w$ is the window size (a hyper-parameter), $t_{i-1} \in \mathbb{R}^{1 \times d}$ is previous tag embedding, $S_{i}$ is embedding of current root. 
These embedding matrix is initialized with small random numbers and trained by back-propagation.

\subsection{Tensor Layer}
In order to capture more interactions between roots, surface words, tags and entity features, we extend the DNN to the tensor neural network.
We use 3-way tensor $ \mathrm{T} \in \mathbb{R}^{h_2 \times h_1 \times h_1}$, where $h_1$ is size of previous layer and $h_2$ is size of tensor layer.
We define the output of a tensor product $h$ via the following vectorized notation.
\begin{equation}
\begin{aligned}
  h = g(e^T \mathrm{T} e + W^3e + b^3)
\end{aligned}
\label{eq:1}
\end{equation}
where $e \in \mathbb{R}^{h_1} $ is output of previous layer, $W^3 \in \mathbb{R}^{h_2 \times h_1} $, $h \in \mathbb{R}^{h_2}$.
Maintaining the full tensor directly leads to parametric explosion.
Here, we use a tensor factorization approach~\cite{pei2014maxmargin} that factorizes each tensor slice as the product of two low-rank matrices, and get the factorized tensor function:
\begin{equation}
\begin{aligned}
  h = g(e^\mathrm{T} P^{[i]} Q^{[i]}e + W^3e + b^3)
\end{aligned}
\label{eq:1}
\end{equation}
where the matrix $P^{[i]} \in \mathbb{R}^{h_1 \times r}$ and $Q^{[i]} \in \mathbb{R}^{r \times h_1}$  are two low rank matrices, and $r$ is number of the factors (a hyper-parameter).
%
% We believe that the tensor layer better captures the interaction of root, word and tag embeddings.
% %
% Experimental results also verify the success of the tensor layer with named entity label embeddings (see more comparisons in Section 4).

\subsection{Tag inference}
There are strong dependencies between the named entity tags in a sentence for the NER. 
In order to capture the tag transitions, we use a transition score $A_{ij}$ ~\cite{Collobert:2011,Zheng:2013} for jumping from one tag $i \in \mathcal{T}$ to another tag $j \in \mathcal{T}$ and an initial scores $A_{0i}$ for starting from the $i^{th}$ tag.
For the input sentence $X$ with a tag sequence $Y$, a sentence-level score can be calculated by the sum of transition and the output of NNs:
\begin{equation}
\begin{aligned}
s(X,Y,\theta) = \sum_{n=1}^{N}(A_{t_{i-1},t_i} + f_{\theta}(t_i|i))
\end{aligned}
\label{eq:1}
\end{equation}
where $f_{\theta}(t_i|i)$ indicates the score output by the network for the $t_{i}$ tag at the $i^{th}$ word.
It should be noted that this model calculates the tag transition score independently from NNs.

One possible way of combining the both tag transitions and neural network outputs is to feed the previous tag embedding to the NNs.
Then, the output of NNs could calculate a transition score given the previous tag embedding, and it can be written as follows:
\begin{equation}
\begin{aligned}
s(X,Y,\theta) = \sum_{n=1}^{N}f_{\theta}(t_i|i,t_{i-1})
\end{aligned}
\label{eq:1}
\end{equation}

At inference time, for a sentence $X$, we can find the best tag path $Y^*$ by maximizing the sentence score. The Viterbi algorithm can be used for this inference.

\section{Experiments}
We conducted several experiments to evaluate our NNs.
One of them is to explore the effects of the word, root and tag embedding plus the tensor layer for MCL NER task, independently.
Another is to show the results of our models after using the pre-trained root and word embeddings.
The last is to compare our models to the state-of-the-art including character embedding-based biLSTM-CRF \cite{N16-1030}.

\subsection{Data-set}
In experiments we used the data from \cite{Tur:2003:SIE:973762.973766} for Turkish and the Kazakh NER corpus (KNC) from \cite{Tolegen:16}.
Both corpus were divided into training (80\%), development (10\%) and test (10\%) set.
The development set is for choosing the hyper-parameters and model selection.
We adopted IOB tagging scheme~\cite{TjongKimSang:2002} for all experiments and used standard conlleval evaluation script\footnote{\scriptsize {www.cnts.ua.ac.be/conll2000/chunking/conlleval.txt}} to report the F-score, precision and recall values.

% \begin{table}[!t]
% \centering
% \label{my-label}
% \caption{ \label{font-table} The hyper-parameters }
% \begin{tabular}{c c c c c c}
% \hline
% Model               & $w$ & $d$  & $h_2$  & $r$  \\
% \hline
% NNs                 &  3   &  50  &  300  &  -   \\
% NNs with tensor       &  3   &  25  &  50   &  3  \\
% \hline
% \end{tabular}
% \end{table}

\begin{table}[]
\caption{Corpus statistics.}
\begin{tabular}{l|lllll|lllll}
\hline
      & \multicolumn{5}{c|}{Kazakh}               & \multicolumn{5}{c}{Turkish}              \\ \hline
      & \#sent. & \#token & \#LOC & \#ORG & \#PER & \#sent. & \#token & \#LOC & \#ORG & \#PER \\ \hline
train & 14457   & 215448  & 5870  & 2065  & 3424  & 22050   & 397062  & 9387  & 7389  & 13080 \\
dev.  & 1807    & 27277   & 785   & 247   & 413   & 2756    & 48990   & 1171  & 869   & 1690  \\
test  & 1807    & 27145   & 731   & 247   & 452   & 2756    & 46785   & 1157  & 925   & 1521  \\ \hline
\end{tabular}
\end{table}

\subsection{Model setup}
A set of experiments were conducted to chose the hyper-parameters and the hyper-parameters are tuned on the development set.
The initial learning rate of AdaGrad is set to 0.01 and the regularization is fixed to $10^{-4}$.
Generally, the number of hidden units has a limited impact on the performance as long as it is large enough.
%
%The final hyper-parameters of our models are listed in Table 2. 
%
Window size $w$ was set to 3, the word, root and tag embedding size was set to 50, number of hidden units was 300 for NNs, and for those NNs with tensor layer, it was set to 50 and its factor size was set to 3.
After finding the best hyper-parameters, we would train final models for all NNs.
After each epoch over the training set, we measured the accuracy of the model on the development set and chose the final model that obtained the highest performance on development set, then use the test set to evaluate the selected model. 
We made several pre-processing to the corpora, namely token and sentence segmentation, lowercasing surface words and the roots were kept in original forms.

\subsection{Results}
We evaluate the following model variations in the experiment: 
i) a baseline neural network, \textit{NN}, which contains a discrete tag transition;
ii) \textit{NN+root} refers to a model that uses root embedding and the discrete tag transition.
iii) \textit{NN+root+tag} is a model that the discrete tag transition in NN is replaced by named entity tag embedding.
iv) \textit{NN+root+tensor} refers to tensor layer-based model with discrete tag transition.
v) models with \textit{+feat} refer to the models use the named entity feature.

% Please add the following required packages to your document preamble:
% \usepackage{multirow}
\begin{table}[]
\caption{Results of the NNs for Kazakh and Turkish (F1-score, \%). Here \textit{root} and \textit{tag} indicate root and tag embeddings; \textit{tensor} means tensor layer; \textit{feat} denotes entity feature vector; \textit{Kaz} - Kazakh and \textit{Tur} - Turkish; \textit{Ov} - Overall.}
\begin{tabular}{c|c|c|c|l|lcllcc|cccllcll}
\hline
L.                   & \# & Models                  & \multicolumn{8}{c|}{Development set}                                   & \multicolumn{8}{c}{Test set}                                                                             \\ \hline
\multirow{8}{*}{Kaz} &    &                         & \multicolumn{3}{c}{LOC}   & \multicolumn{3}{c}{ORG}   & PER   & Ov    & LOC            & ORG            & \multicolumn{3}{c}{PER}            & \multicolumn{3}{c}{Ov}             \\ \cline{2-19} 
                     & 1  & NN                      & \multicolumn{3}{c}{86.69} & \multicolumn{3}{c}{68.95} & 68.57 & 78.66 & 86.32          & 69.51          & \multicolumn{3}{c}{64.78}          & \multicolumn{3}{c}{76.89}          \\
                     & 2  & NN+root                 & \multicolumn{3}{c}{87.48} & \multicolumn{3}{c}{70.23} & 75.66 & 81.20 & 87.74          & 72.53          & \multicolumn{3}{c}{\textbf{75.25}} & \multicolumn{3}{c}{\textbf{81.36}} \\
                     & 3  & NN+root+tag             & \multicolumn{3}{c}{88.85} & \multicolumn{3}{c}{67.69} & 79.68 & 82.81 & 87.65          & 73.75          & \multicolumn{3}{c}{76.13}          & \multicolumn{3}{c}{\textbf{81.86}} \\
                     & 4  & NN+root+tensor          & \multicolumn{3}{c}{89.56} & \multicolumn{3}{c}{72.54} & 81.07 & 84.22 & 88.51          & \textbf{75.79} & \multicolumn{3}{c}{77.32}          & \multicolumn{3}{c}{\textbf{82.83}} \\ \cline{2-19} 
                     & 5  & NN+root+feat            & \multicolumn{3}{c}{93.48} & \multicolumn{3}{c}{78.35} & 91.59 & 90.40 & 92.48          & 78.90          & \multicolumn{3}{c}{90.75}          & \multicolumn{3}{c}{89.54}          \\
                     & 6  & NN+root+tensor+feat     & \multicolumn{3}{c}{93.78} & \multicolumn{3}{c}{81.48} & 90.91 & 90.87 & 92.22          & \textbf{81.57} & \multicolumn{3}{c}{91.27}          & \multicolumn{3}{c}{90.11}          \\
                     & 7  & NN+root+tag+tensor+feat & \multicolumn{3}{c}{93.65} & \multicolumn{3}{c}{81.28} & 92.42 & 91.27 & \textbf{92.96} & 78.89          & \multicolumn{3}{c}{\textbf{91.70}} & \multicolumn{3}{c}{\textbf{90.28}} \\ \hline
\multirow{7}{*}{Tur} & 8  & NN                      & \multicolumn{3}{c|}{85.06} & \multicolumn{3}{c}{74.70} & 81.11 & 80.86 & 83.17          & 76.26          & \multicolumn{3}{c}{80.55}          & \multicolumn{3}{c}{80.29}          \\
                     & 9  & NN+root                 & \multicolumn{3}{c}{87.38} & \multicolumn{3}{c}{77.13} & 84.78 & 83.78 & 85.78          & 78.66          & \multicolumn{3}{c}{84.03}          & \multicolumn{3}{c}{\textbf{83.17}} \\
                     & 10 & NN+root+tag             & \multicolumn{3}{c}{90.70} & \multicolumn{3}{c}{84.93} & 86.67 & 87.53 & \textbf{90.02} & \textbf{86.14} & \multicolumn{3}{c}{85.95}          & \multicolumn{3}{c}{\textbf{87.31}} \\
                     & 11 & NN+root+tensor          & \multicolumn{3}{c}{92.43} & \multicolumn{3}{c}{86.45} & 89.63 & 89.78 & 90.50          & 87.14          & \multicolumn{3}{c}{\textbf{90.00}} & \multicolumn{3}{c}{\textbf{89.42}} \\ \cline{2-19} 
                     & 12 & NN+root+feat            & \multicolumn{3}{c}{91.54} & \multicolumn{3}{c}{89.04} & 91.62 & 91.01 & 90.27          & 89.50          & \multicolumn{3}{c}{91.95}          & \multicolumn{3}{c}{90.78}          \\
                     & 13 & NN+root+tensor+feat     & \multicolumn{3}{c|}{93.60} & \multicolumn{3}{c}{88.88} & 92.23 & 91.88 & 92.05          & 89.35          & \multicolumn{3}{c}{92.01}          & \multicolumn{3}{c}{91.34}          \\
                     & 14 & NN+root+tag+tensor+feat & \multicolumn{3}{c}{91.77} & \multicolumn{3}{c}{89.72} & 92.23 & 91.44 & 92.80          & 88.45          & \multicolumn{3}{c}{91.91}          & \multicolumn{3}{c}{\textbf{91.39}} \\ \hline
\end{tabular}
\end{table}

Table 3 summaries the results for Kazakh and Turkish.
Rows (1-4, 8-11) are given to compare the root, tag embedding and tensor layer independently.
Rows (5-7, 12-14) shows the effect of entity features.
As shown, when only use the surface word forms, the \textit{NN} gives 76.89\% overall F1-score for Kazakh.
The \textit{NN} gives low F1-scores of 64.78\% and 69.51\% for PER and ORG respectively.
There are mainly two reasons for this:  i) the number of person and organization names are less than location (Table 2), and ii) compared to other entities, the length of organization name is much longer, it also has ambiguous words with people names\footnote{It often appears when the organization name is given after someone's name.}.
For Turkish, \textit{NN} yields 80.29\% overall F1.

It is evident from (row 2, 9) that \textit{NN+root} is improved significantly in all terms after using the root embedding.
There are 4.47\% and 2.88\% improvements in overall F1 for Kazakh and Turkish compare to \textit{NN}.
More precisely, using root embedding, \textit{NN+root} gives 10.47\%, 3.02\% and 1.42\% improvements for Kazakh PER, ORG, LOC entities, respectively.
The result for Turkish also follows the pattern.
Row (3,10) shows the effect of replacing the discrete tag transition with named entity tag embedding.
We could observe that \textit{NN+root+tag} yields overall F1-scores of 81.86\% and 87.31\% for Kazakh and Turkish.
Compared to \textit{NN+root}, the model with entity tag embedding has a significant improvement for Turkish with 4.14\% in overall F1. 
For two languages, the model performances are boosted by using tensor transformation; it shows that the tensor layer could capture the more interactions between root and word vectors.
Using the entity features, \textit{NN+root+feat} give a significant improvement   for Kazakh (from 81.36 to 89.54\% ) and Turkish (from 83.17 to 90.78\%).
The best result for Kazakh is 90.28\% F1-score that is obtained by using tensor transformation with tag embeddings and entity features.

%%%%%%%%%%%%%%%%%%%%%%
% \begin{figure}[!ht]
% \centering
% \includegraphics[
%   page=1,
%   width=270,
%   height=180,
%   keepaspectratio,
%   trim=0 0 0 0,]{training_curve.png}
%   \caption{The training accuracy of NNs.}
% \end{figure}

% Figure 2 shows the accuracy of the training process at the first 50 training epochs.
% %
% It can be seen that the NNs without using root embedding converge slowly compared to that of using root embedding.
% %
% The use of tag embedding and tensor layer also have an impact on reducing the training time.
% %
% It shows that the models \textit{NN+tag} and \textit{NN+tensor} converge quicker than the model \textit{NN} using only surface word form.
% %
% In practice, we notice that using root embedding could save a lot of time for training and it also provides the significant growths on performance (see Table 3).
% %
% It indicates that root embedding is vital for MCL and it could lead to significant improvements for many NLP task of MCL in term of reducing the training time and improving model performance.
%%%%%%%%%%%%%%%

We compare our NNs with exiting CRF-based NER system~\cite{Tolegen:16} and other state-of-the-art models.
According to the recent studies for NER \cite{N16-1030,P16-1101,W18-5605}, the current cutting-edge deep learning models for sequence labeling problem is bi-directional LSTM with CRF layer.
On the one hand, we trained such state-of-the-art NER model for Kazakh language for making comparisons.
On the other, It is also worth to see how does a character-based model perform well for agglutinative languages.
Because the character-based approaches seem to be well suited for agglutinative nature of the languages and it can serve as a stronger baseline than CRF.
For those biLSTM-based models, we set hyper-parameters are comparable with those models yield the state-of-the-art results for English \cite{N16-1030,P16-1101}.
The word and character embeddings are set to $300$ and $100$, respectively. 
The hidden unit of LSTM for both character and word are set to $300$.
The dropout is set to 0.5 and use "Adam" updating strategy for learning model parameters.
It should be note that the form of entities in Kazakh always starts with capital letter, and the data set used for all biLSTM-based models are not converted to lowercase, which could lead a positive effect for recognition. 
For a fair comparison, the following NER models are trained on the same training, development and test set.
Table 4 shows the comparison of our NNs with state-of-the-art for Kazakh.

\begin{table}[]
\caption{Comparison of our NNs and state-of-the-art}
\centering
\begin{tabular}{c|cccc}
\hline
Models                                       & LOC            & ORG            & PER            & Overall        \\ \hline
CRF \cite{Tolegen:16} & 91.71          & \textbf{83.40} & 90.06          & \textbf{89.81} \\
biLSTM+dropout                               & 85.84          & 68.91          & 72.75          & 78.76          \\
biLSTM-CRF+dropout                           & 86.52          & 69.57          & 75.79          & 80.28          \\
biLSTM-CRF+Characters+dropout                & 90.43          & 76.10          & 85.88          & \textbf{86.45} \\ \hline
NN+root+feat                                 & 92.48          & 78.90          & 90.75          & 89.54          \\
NN+root+tensor+feat                          & 92.22          & 81.57          & 91.27          & 90.11          \\
NN+root+tag+tensor+feat                      & \textbf{92.96} & 78.89          & 91.70          & \textbf{90.28} \\ \hline
NN+root+feat*                                & 91.74          & 81.00          & 90.99          & 89.70          \\
NN+root+tensor+feat*                         & 92.91          & 81.76          & 91.09          & 90.40          \\
NN+root+tag+tensor+feat*                     & 91.33          & \textbf{81.88} & \textbf{92.00} & \textbf{90.49} \\ \hline
\end{tabular}
\end{table}

The CRF-based system~\cite{Tolegen:16} achieved an F1-score of 89.81\% using all features with their well-designed feature template.
The biLSTM-CRF with character embedding yields 86.45\% F1-score which is better than the result of the model without using characters.
It can be seen,  the significant improvement about 6\% in overall F1-score was gained after using character embeddings.
It indicates that character-based model fits the nature of the MCL.
We initialized the root and word embedding by using pre-trained embeddings.
The skip-gram model of \emph{word2vec}\footnote{\scriptsize {https://code.google.com/p/word2vec/}.} ~\cite{Mikolov:13} is used to train root and word vectors on large Kazakh news articles and Wikipedia texts \footnote{\scriptsize {In order to reduce dictionary size of root and surface word, we did some pre-processing namely, lowercasing and word stemming by morphological analyzer and disambiguator.}}. 
Table 4 also shows the results after pre-training the root and word embedding marked with symbol *. 
As shown, the pre-trained root and word representations have a minor effect on the overall F1-score of NN models.
Especially for organization names, the pre-trained embeddings have positive effects.
The \textit{NN+root+feat*} and the \textit{NN+root+tag+tensor+feat*} models achieve around 2\% improvement for organization F1-score compared to those of the models without using the per-trained embeddings (the former's is form 78.90\% to 81.00\% and the latter's is from 78.89\% to 81.88\%).
Overall, our NN outperforms the CRF-based system and other state-of-the-art (biLSTM-CRF-character+dropout), and the best NN yields an F1 of 90.49\%, a new state-of-the-art for Kazakh NER.

To show the effect of word embeddings after the model training. 
We calculated the ten nearest neighbors of a few randomly chosen query words (first row). 
Their distances were measured by the cosine similarity.
As given in Table 5, the nearest neighbors in three columns are related to their named entity labels: all location, person and organization names are listed in the first, second and third column, respectively.
Compared to CRF, instead of using discrete features, the NNs project root, words into a vector space, which could group similar words by their meaning and the NNs has non-linear transformations to extract higher-level features.
In this way,  the NNs may reduce the effects of data sparsity problems of MCL.

\label{sect:pdf}
\begin{table}[h]
\begin{center}
\caption{ \label{font-table} Example words in Kazakh and their 10 closest neighbors. Here, we used the Latin alphabet to write Kazakh words for convenience.}
\begin{tabular}{c|c|c}
\hline
\emph{ Kazakhstan} (Location)  &  \emph{Meirambek} (Person)&  \emph{KazMunayGas} (Organization) \\
\hline
Kiev     & Oteshev & Nurmukasan  \\
Sheshenstandagy     & Klinton & TsesnaBank  \\
Kyzylorda     & Shokievtin  & Euroodaktyn\\
Angliada     & Dagradorzh   & Atletikony  \\
Burabai       & Tarantinonyn   & Bayern  \\
Iran       & Nikliochenko   & Euroodakka\\
Singapore       & Luis   & CenterCredittin  \\
Neva       & Monhes   & Juventus  \\
London       & Fernades   & Aldaraspan  \\
Romania       & Fog   & Liverpool  \\
\hline
\end{tabular}
\end{center}
\end{table}

\section{Conclusions}
We presented several neural networks for NER of MCLs.
The key aspects of our model for MCL are to utilize different embeddings and layer,  namely, i) root embedding, ii) entity tag embedding and iii) the tensor layer.
The effects of those aspects are investigated individually.
The use of root embedding leads to a significant result on MCLs' NER.
The other two also gives positive effects.
For Kazakh, the proposed NNs outperform the CRF-based NER system and other state-of-the-art including character-based biLSTM-CRF model.
The comparisons showed that character embedding is vital to MCL's NER.
The experimental results indicate that the proposed NNs can be potentially applied to other morphologically complex languages.

\section*{Acknowledgments}
The work was funded by the Committee of Science of Ministry of Education and Science of the Republic of Kazakhstan under the grant AP09259324.
%and the grant of num. AP05131207 ``Development of technologies for multilingual automatic speech recognition using deep neural networks ''.

%
% ---- Bibliography ----
%
% BibTeX users should specify bibliography style 'splncs04'.
% References will then be sorted and formatted in the correct style.
%

\bibliographystyle{splncs04}
\bibliography{ref}

\begin{thebibliography}{10}
\providecommand{\url}[1]{\texttt{#1}}
\providecommand{\urlprefix}{URL }
\providecommand{\doi}[1]{https://doi.org/#1}

\bibitem{articleOrken:2018}
Baba~Ali, B., Wójcik, W., Orken, M., Turdalyuly, M., Mekebayev, N.: Speech
  recognizer-based non-uniform spectral compression for robust mfcc feature
  extraction. Przeglad Elektrotechniczny  \textbf{94},  90--93 (06 2018).
  \doi{10.15199/48.2018.06.17}

\bibitem{DBLP:journals/corr/BahdanauCB14}
Bahdanau, D., Cho, K., Bengio, Y.: Neural machine translation by jointly
  learning to align and translate. CoRR  \textbf{abs/1409.0473} (2014)

\bibitem{Bengio:2003}
Bengio, Y., Ducharme, R., Vincent, P., Janvin, C.: A neural probabilistic
  language model. J. Mach. Learn. Res.  \textbf{3},  1137--1155 (Mar 2003)

\bibitem{Chieu:2003:NER}
Chieu, H.L., Ng, H.T.: Named entity recognition with a maximum entropy
  approach. In: Proceedings of the Seventh Conference on Natural Language
  Learning at HLT-NAACL 2003 - Volume 4. pp. 160--163. CONLL '03, Association
  for Computational Linguistics, Stroudsburg, PA, USA (2003)

\bibitem{Collobert:2011}
Collobert, R., Weston, J., Bottou, L., Karlen, M., Kavukcuoglu, K., Kuksa, P.:
  Natural language processing (almost) from scratch. J. Mach. Learn. Res.
  \textbf{12},  2493--2537 (Nov 2011)

\bibitem{Florian:2003:NER}
Florian, R., Ittycheriah, A., Jing, H., Zhang, T.: Named entity recognition
  through classifier combination. In: Proceedings of the Seventh Conference on
  Natural Language Learning at HLT-NAACL 2003 - Volume 4. pp. 168--171. CONLL
  '03, Association for Computational Linguistics, Stroudsburg, PA, USA (2003)

\bibitem{Graves06connectionisttemporal}
Graves, A., Fernández, S., Gomez, F.: Connectionist temporal classification:
  Labelling unsegmented sequence data with recurrent neural networks. In: In
  Proceedings of the International Conference on Machine Learning, ICML 2006.
  pp. 369--376 (2006)

\bibitem{He:NIPS:2016:6469}
He, D., Xia, Y., Qin, T., Wang, L., Yu, N., Liu, T., Ma, W.Y.: Dual learning
  for machine translation. In: Lee, D.D., Sugiyama, M., Luxburg, U.V., Guyon,
  I., Garnett, R. (eds.) Advances in Neural Information Processing Systems 29,
  pp. 820--828. Curran Associates, Inc. (2016)

\bibitem{Klein:2003:NER}
Klein, D., Smarr, J., Nguyen, H., Manning, C.D.: Named entity recognition with
  character-level models. In: Proceedings of the Seventh Conference on Natural
  Language Learning at HLT-NAACL 2003 - Volume 4. pp. 180--183. CONLL '03,
  Association for Computational Linguistics, Stroudsburg, PA, USA (2003)

\bibitem{kuru-can-yuret:2016:COLING}
Kuru, O., Can, O.A., Yuret, D.: Charner: Character-level named entity
  recognition. In: Proceedings of COLING 2016, the 26th International
  Conference on Computational Linguistics: Technical Papers. pp. 911--921. The
  COLING 2016 Organizing Committee, Osaka, Japan (December 2016)

\bibitem{Lafferty:2001}
Lafferty, J.D., McCallum, A., Pereira, F.C.N.: Conditional random fields:
  Probabilistic models for segmenting and labeling sequence data. In:
  Proceedings of the Eighteenth International Conference on Machine Learning.
  pp. 282--289. ICML '01, Morgan Kaufmann Publishers Inc., San Francisco, CA,
  USA (2001)

\bibitem{N16-1030}
Lample, G., Ballesteros, M., Subramanian, S., Kawakami, K., Dyer, C.: Neural
  architectures for named entity recognition. In: Proceedings of the 2016
  Conference of the North American Chapter of the Association for Computational
  Linguistics: Human Language Technologies. pp. 260--270. Association for
  Computational Linguistics (2016)

\bibitem{ling-EtAl:2015:EMNLP2}
Ling, W., Dyer, C., Black, A.W., Trancoso, I., Fermandez, R., Amir, S., Marujo,
  L., Luis, T.: Finding function in form: Compositional character models for
  open vocabulary word representation. In: Proceedings of the 2015 Conference
  on Empirical Methods in Natural Language Processing. pp. 1520--1530.
  Association for Computational Linguistics, Lisbon, Portugal (September 2015)

\bibitem{P16-1101}
Ma, X., Hovy, E.: End-to-end sequence labeling via bi-directional
  lstm-cnns-crf. In: Proceedings of the 54th Annual Meeting of the Association
  for Computational Linguistics (Volume 1: Long Papers). pp. 1064--1074.
  Association for Computational Linguistics (2016). \doi{10.18653/v1/P16-1101},
  \url{http://aclweb.org/anthology/P16-1101}

\bibitem{10.1080/23311916.2020.1727168}
Mamyrbayev, O., Toleu, A., Tolegen, G., Mekebayev, N.: Neural architectures for
  gender detection and speaker identification. Cogent Engineering
  \textbf{7}(1) (2020). \doi{10.1080/23311916.2020.1727168},
  \url{http://doi.org/10.1080/23311916.2020.1727168}

\bibitem{orken2018}
Mamyrbayev, O., Turdalyuly, M., Mekebayev, N., Mukhsina, K., Alimhan, K.,
  BabaAli, B., Nabieva, G., Duisenbayeva, A., Akhmetov, B.: Continuous speech
  recognition of kazakh language. ITM Web of Conferences  \textbf{24},  01012
  (01 2019). \doi{10.1051/itmconf/20192401012}

\bibitem{McCallum:2003:NER}
McCallum, A., Li, W.: Early results for named entity recognition with
  conditional random fields, feature induction and web-enhanced lexicons. In:
  Proceedings of the Seventh Conference on Natural Language Learning at
  HLT-NAACL 2003 - Volume 4. pp. 188--191. CONLL '03, Association for
  Computational Linguistics, Stroudsburg, PA, USA (2003)

\bibitem{Mikolov:13}
Mikolov, T., Chen, K., Corrado, G., Dean, J.: Efficient estimation of word
  representations in vector space. CoRR  \textbf{abs/1301.3781} (2013)

\bibitem{pei2014maxmargin}
Pei, W., Ge, T., Chang, B.: Max-margin tensor neural network for chinese word
  segmentation. In: Proceedings of the 52nd Annual Meeting of the Association
  for Computational Linguistics (Volume 1: Long Papers). pp. 293--303.
  Association for Computational Linguistics, Baltimore, Maryland (June 2014)

\bibitem{DBLP:conf/coling/SekerE12}
Seker, G.A., Eryigit, G.: Initial explorations on using crfs for turkish named
  entity recognition. In: {COLING} 2012, 24th International Conference on
  Computational Linguistics, Proceedings of the Conference: Technical Papers,
  8-15 December 2012, Mumbai, India. pp. 2459--2474 (2012)

\bibitem{TjongKimSang:2002}
Tjong Kim~Sang, E.F.: Introduction to the conll-2002 shared task:
  Language-independent named entity recognition. In: Proceedings of the 6th
  Conference on Natural Language Learning - Volume 20. pp.~1--4. COLING-02,
  Association for Computational Linguistics, Stroudsburg, PA, USA (2002)

\bibitem{TjongKimSang:2003}
Tjong Kim~Sang, E.F., De~Meulder, F.: Introduction to the conll-2003 shared
  task: Language-independent named entity recognition. In: Proceedings of the
  Seventh Conference on Natural Language Learning at HLT-NAACL 2003 - Volume 4.
  pp. 142--147. CONLL '03, Association for Computational Linguistics,
  Stroudsburg, PA, USA (2003)

\bibitem{konvens:17_tkachenko12o}
Tkachenko, M., Simanovsky, A.: Named entity recognition: {Exploring} features.
  In: Jancsary, J. (ed.) Proceedings of KONVENS 2012. pp. 118--127. \"{O}GAI
  (September 2012), main track: oral presentations

\bibitem{Tolegen:16}
Tolegen, G., Toleu, A., Zheng, X.: Named entity recognition for kazakh using
  conditional random fields. In: Proceedings of the 4-th International
  Conference on Computer Processing of Turkic Languages TurkLang 2016. pp.
  118--127. Izvestija KGTU im.I.Razzakova (2016)

\bibitem{8880244}
{Toleu}, A., {Tolegen}, G., {Mussabayev}, R.: Comparison of various approaches
  for dependency parsing. In: 2019 15th International Asian School-Seminar
  Optimization Problems of Complex Systems (OPCS). pp. 192--195 (2019)

\bibitem{toleu-tolegen-makazhanov:2017:Short}
Toleu, A., Tolegen, G., Makazhanov, A.: Character-aware neural morphological
  disambiguation. In: Proceedings of the 55th Annual Meeting of the Association
  for Computational Linguistics (Volume 2: Short Papers). pp. 666--671.
  Association for Computational Linguistics, Vancouver, Canada (July 2017)

\bibitem{Tur:2003:SIE:973762.973766}
T\"{u}r, G., Hakkani-t\"{u}r, D., Oflazer, K.: A statistical information
  extraction system for turkish. Nat. Lang. Eng.  \textbf{9}(2),  181--210 (Jun
  2003)

\bibitem{wieting-EtAl:2016:EMNLP2016}
Wieting, J., Bansal, M., Gimpel, K., Livescu, K.: Charagram: Embedding words
  and sentences via character n-grams. In: Proceedings of the 2016 Conference
  on Empirical Methods in Natural Language Processing. pp. 1504--1515.
  Association for Computational Linguistics, Austin, Texas (November 2016)

\bibitem{Yeniterzi:2011:EMT}
Yeniterzi, R.: Exploiting morphology in turkish named entity recognition
  system. In: Proceedings of the ACL 2011 Student Session. pp. 105--110. HLT-SS
  '11, Association for Computational Linguistics, Stroudsburg, PA, USA (2011)

\bibitem{W18-5605}
Zhai, Z., Nguyen, D.Q., Verspoor, K.: Comparing cnn and lstm character-level
  embeddings in bilstm-crf models for chemical and disease named entity
  recognition. In: Proceedings of the Ninth International Workshop on Health
  Text Mining and Information Analysis. pp. 38--43. Association for
  Computational Linguistics (2018)

\bibitem{Zheng:2013}
Zheng, X., Chen, H., Xu, T.: Deep learning for {Chinese} word segmentation and
  {POS} tagging. In: Proceedings of the 2013 Conference on Empirical Methods in
  Natural Language Processing. pp. 647--657. Association for Computational
  Linguistics, Seattle, Washington, USA (October 2013)

\end{thebibliography}
\end{document}